\documentclass[pre,aps]{revtex4}
\usepackage{graphicx}

\begin{document}

\title{Universal low-temperature properties of frustrated classical spin
chain near the ferromagnet-helimagnet transition point}
\author{D.~V.~Dmitriev}
\email{dmitriev@deom.chph.ras.ru}
\author{V.~Ya.~Krivnov}
\affiliation{Joint Institute of Chemical Physics, RAS, Kosygin
str. 4, 119334, Moscow, Russia.}
\date{}

\begin{abstract}
The thermodynamics of the classical frustrated spin chain near the
transition point between the ferromagnetic and the helical phases
is studied. The calculation of the partition and spin correlation
functions at low temperature limit is reduced to the quantum
mechanical problem of a particle in potential well. It is shown
that the thermodynamic quantities are universal functions of the
temperature normalized by the chiral domain wall energy. The
obtained behavior of the static structure factor indicates that
the short-range helical-type correlations existing at low
temperatures on the helical side of the transition point disappear
at some critical temperature, defining the Lifshitz point. It is
also shown that the low-temperature susceptibility in the helical
phase near the transition point has a maximum at some temperature.
Such behavior is in agreement with that observed in several
materials described by the quantum $s=1/2$ version of this model.
\end{abstract}

\maketitle

\section{Introduction}

Strongly frustrated low-dimensional magnets attracted much attention
last years \cite{mikeskabook}. A very interesting class of such
compounds is edge-sharing chains where $CuO_{4}$ plaquets are
coupled by their edges \cite{Mizuno,Hase,Malek,Nitzsche}. An
important feature of the edge-sharing chains is that the
nearest-neighbor (NN) interaction $J_1$ between $Cu$ spins is
ferromagnetic while the next-nearest-neighbor (NNN) interaction
$J_2$ is antiferromagnetic. The competition between them leads to
the frustration. A minimal model describing the magnetic properties
of these cuprates is so-called F-AF spin chain model the Hamiltonian
of which is
\begin{equation}
H=J_{1}\sum \mathbf{S}_{n}\cdot \mathbf{S}_{n+1}+J_{2}\sum
\mathbf{S} _{n}\cdot \mathbf{S}_{n+2}  \label{H}
\end{equation}
where $J_{1}<0$ and $J_{2}>0$.

The ground state phase diagram of the $s=1/2$ F-AF model has been
studied extensively last years
\cite{Chubukov,Itoi,Dmitriev,Vekua,Lu,Richter,Kecke}. The ground
state of this model is governed by the frustration parameter $\alpha
=J_2/\left\vert J_1\right\vert $. At $0<\alpha<1/4$ the ground state
is fully ferromagnetic. At $\alpha>1/4$ the incommensurate singlet
phase with short-range helical spin correlations is realized. At
$\alpha=1/4$ the quantum phase transition occurs. Remarkably, this
transition point does not depend on a spin value, including the
classical limit $s\to\infty$.

However, the influence of the frustration on low temperature
thermodynamics is less studied, especially in the vicinity of the
ferromagnet-helimagnet transition point. It is of a particular
importance to study this problem, because there are several
edge-sharing cuprates with $\alpha \simeq 1/4$ which are of
special interest \cite{Volkova}.

Unfortunately, at present the low-temperature thermodynamics of
quantum $s=1/2$ model (\ref{H}) at $\alpha \neq 0$ can be studied
only either by using of numerical calculations of finite chains or
by approximate methods. On the other hand, the classical version of
model (\ref{H}) can be studied exactly at $T\to 0$. It is expected
that the main qualitative features of the quantum low-temperature
thermodynamics can be reproduced correctly in the framework of the
classical model. Besides, the classical limit can be used as a
starting point to study quantum effects. Therefore, the study of
classical model (\ref{H}) is useful for the understanding of the
low-temperature properties of the quantum F-AF chains.

At zero temperature classical model (\ref{H}) has magnetic long
range-order (LRO) for all values of $\alpha $: the ferromagnetic LRO
at $\alpha \leq 1/4$ and the helical-type LRO with the wave number
$Q=\cos^{-1}(1/4\alpha)$ for $\alpha>1/4$. It should be noticed that
both the time-reversal and the parity symmetries are broken in the
helical ordered phase. The helical ordered state possesses a
two-fold discrete chiral degeneracy (in addition to the usual
spin-rotational degeneracy) corresponding to clockwise and
counter-clockwise turn of spins. The chiral order parameter is
defined by the chirality vector $\vec{K}_{n}=[\vec{S}_{n}\times
\vec{S}_{n+1}]$. At finite temperature the LRO is destroyed by
thermal fluctuations, the thermodynamic quantities have singular
behavior at $T\to 0$ with the corresponding critical exponents
depending on $\alpha $. In Ref.\cite{DK102} we studied classical
model (\ref{H}) at $\alpha =1/4$, i.e. exactly at the
ferromagnet-helimagnet transition point. It was shown that the
correlation length $l_{c}$ behaves as $T^{-1/3}$ and zero-field
susceptibility $\chi $ diverges as $T^{-4/3}$ in contrast with the
1D Heisenberg ferromagnet (HF) ($\alpha =0$) where $l_{c}\sim
T^{-1}$ and $\chi \sim T^{-2}$ \cite{Fisher}.

In this paper we focus on the low-temperature thermodynamics of the
F-AF classical model in the vicinity of the ferromagnet-helimagnet
transition point. As it will be shown, in the low-temperature limit
the thermodynamic quantities become the universal functions of the
scaled temperature $t=T/\gamma ^{3/2}$, where the parameter $\gamma
=(4\alpha -1)$ characterizes the deviation from the transition point
and is assumed to be small. The calculation of the thermodynamic
quantities in this limit reduces to the solution of the
Schr\"{o}dinger equation for the quantum particle in an anharmonic
potential.

It is worth noting that the thermodynamics of classical spin model
(\ref{H}) has been studied before \cite{Harada0,Harada} using the
transfer-matrix method with the subsequent solution of the
corresponding integral equations by Gaussian integration formulas.
This method gives reliable results for the moderate temperatures and
values of $\gamma$. However, this procedure fails for low
temperatures because of appearing of an artificial gap in the
excitation spectrum \cite{Pandit}. Our approach gives exact
asymptotics of the thermodynamic quantities at $T\to 0$, $\gamma \to
0$ and these results are complementary to those obtained in
\cite{Harada0,Harada}.

The paper is organized as follows. In Section II the continuum
version of the model is introduced. The problem of the calculation
of the partition function is reduced to the solution of the
Schr\"{o}dinger equation for a particle in the potential well. In
Section III the spin correlation function is obtained and the
$k$-dependence of the static structure factor is studied. It is
shown that the low temperature thermodynamics is the universal
function of the scaling variable. The phase diagram of the model is
constructed and the location of the Lifshitz point is determined.
The behavior of the chiral correlation function is also studied. In
Section IV the summary of the obtained results is given and their
relation to early studies of the considered model
\cite{Harada0,Harada} is discussed. In the Appendix the derivation
of the energy of chiral domain wall is given.

\section{Partition function}

In the vicinity of the transition point it is convenient to rewrite
Hamiltonian (\ref{H}) in the form \cite{DK10}:
\begin{equation}
H=\frac{1}{8}\sum \left[
(\mathbf{S}_{n+1}-2\mathbf{S}_{n}+\mathbf{S} _{n-1})^{2}-\gamma
(\mathbf{S}_{n+1}-\mathbf{S}_{n-1})^{2}\right] \label{H14}
\end{equation}
where $\gamma =(4\alpha -1)$.

In Eq.(\ref{H14}) we put $|J_{1}|=1$ and omit unessential constant.
In the classical approximation the spin operators $\mathbf{S}_{n}$
are replaced by the classical unit vectors $\vec{S}_{n}$. We will
investigate mainly the case $\gamma >0$ corresponding to the helical
side of the ground state phase diagram. In the limit $T\to 0$ the
thermal fluctuations are weak and at $\gamma\to 0$ the period of the
helix is long, so that neighbor spins are directed almost parallel
to each other. Therefore, we can use the continuum approach
replacing $\vec{S}_{n}$ by the classical unit vector field $\vec{s}
(x)$ with slowly varying orientations, so that
\begin{equation}
\vec{S}_{n+1}-\vec{S}_{n-1}\approx 2\frac{\partial
\vec{s}(x_{n})}{\partial x }  \label{c1}
\end{equation}
and
\begin{equation}
\vec{S}_{n+1}-2\vec{S}_{n}+\vec{S}_{n-1}\approx \frac{\partial
^{2}\vec{s} (x_{n})}{\partial x^{2}}  \label{c2}
\end{equation}
where the lattice constant is chosen as unit length.

Using Eqs.(\ref{c1}) and (\ref{c2}) Hamiltonian (\ref{H14}) goes
over into the energy functional of the vector field $\vec{s}(x)$:
\begin{equation}
E=\frac{1}{8}\int \mathrm{d}x\left[ \left( \frac{\partial ^{2}\vec{s}}{
\partial x^{2}}\right) ^{2}-4\gamma \left( \frac{\partial \vec{s}}{\partial
x }\right) ^{2}\right]  \label{Etr}
\end{equation}

This energy functional is a starting point of the investigations
of model ( \ref{H}) in the vicinity of the transition point
$\alpha =1/4$. The partition function is a functional integral
over all configurations of the vector field on a ring of length
$L$
\begin{equation}
Z=\int D\vec{s}(x)\exp \left\{ -\frac{1}{8T}\int_{0}^{L}dx\left[
\left( \frac{\partial ^{2}\vec{s}}{\partial x^{2}}\right)
^{2}-4\gamma \left( \frac{
\partial \vec{s}}{\partial x}\right) ^{2}\right] \right\}  \label{Z0}
\end{equation}

It is useful to scale the spatial variable as
\begin{equation}
\xi =2T^{1/3}x  \label{xi}
\end{equation}

Then, the partition function takes the dimensionless form
\begin{equation}
Z=\int D\vec{s}(\xi )\exp \left\{ -\int_{0}^{\lambda }d\xi \left[
\left( \frac{d^{2}\vec{s}}{d\xi ^{2}}\right) ^{2}-b \left(
\frac{d\vec{s}}{d\xi }\right) ^{2}\right] \right\}  \label{Zxi}
\end{equation}
where
\begin{equation}
b =\frac{\gamma }{T^{2/3}}  \label{gamma}
\end{equation}
and the rescaled system length is $\lambda =2T^{1/3}L$.

Since the integration in the partition function occurs over all
possible spin configurations, we are free to choose any local
coordinate system. It is convenient to choose it so that the $z$
axis at the point $\xi $ is directed along the spin vector
$\vec{s}(\xi )$, so that the spin vector $ \vec{s}(\xi )=(0,0,1)$.

Let us introduce a new vector field
\begin{equation}
\vec{q}(\xi )=\frac{d\vec{s}}{d\xi }=(q_{x},q_{y},q_{z})  \label{q3d}
\end{equation}

Then, the constraint $\vec{s}^{2}(\xi )=1$ converts to the
relations for $ \vec{q}(\xi )$:
\begin{eqnarray}
q_{z} &=&0  \nonumber \\
q_{z}^{\prime } &=&-q_{x}^{2}-q_{y}^{2}  \label{constrain}
\end{eqnarray}
where the prime denotes the space derivative $d/d\xi$.

Then, the first term of the Hamilton function in Eq.(\ref{Etr})
transforms to
\begin{equation}
\left( \frac{d^{2}\vec{s}}{d\xi ^{2}}\right) ^{2}=\left(
\frac{d\vec{q}}{ d\xi }\right) ^{2}=q_{x}^{\prime 2}+q_{y}^{\prime
2}+(q_{x}^{2}+q_{y}^{2})^{2}  \label{sq3d}
\end{equation}

Here we see that the constraint $\vec{s}^{2}=1$ effectively
eliminates the $ q_z$ component from the Hamilton function.
Therefore, henceforth we deal with the $q_{x}$ and $q_{y}$
components of the vector field $\vec{q}$ only, and we denote a
two-component vector field by $\mathbf{q}(\xi )=(q_{x},q_{y}) $.

The partition function in terms of $\mathbf{q}(\xi )$ takes the
form:
\begin{equation}
Z=\int D\mathbf{q}(\xi )\exp \left\{ -\int_{0}^{\lambda }d\xi
\left( \mathbf{ q}^{\prime 2}+\mathbf{q}^{4}-b
\mathbf{q}^{2}\right) \right\} \label{zq3d}
\end{equation}

If we replace $\xi $ by an imaginary time $\xi \to it$ then
partition function (\ref{zq3d}) takes the form of a path integral
of a quantum particle in a potential
$U(\mathbf{q})=\mathbf{q}^{4}-b \mathbf{q}^{2}$. In other words
$Z$ in Eq.(\ref{zq3d}) is the partition function of the quantum
system at `temperature' $1/\lambda $ described by the Hamiltonian
\begin{equation}
\hat{H}_{0}=-\frac{1}{4}\Delta +\mathbf{q}^{4}-b \mathbf{q}^{2}
\label{H0}
\end{equation}
where $\Delta =\partial _{x}^{2}+\partial _{y}^{2}$ is
two-dimensional Laplace operator. Hamiltonian (\ref{H0}) commutes
with the $z$-component of the angular momentum $\hat{l}_{z}$ and
eigenstates $\psi (\mathbf{q})$ of the corresponding
Schr\"{o}dinger equation are divided into subspaces of azimuthal
quantum numbers $l_{z}=0,\pm 1,\pm 2\ldots $
\begin{equation}
\hat{H}_{0}\psi _{n,l_{z}}=\varepsilon _{n,l_{z}}\psi _{n,l_{z}}
\label{Hpsi}
\end{equation}

\begin{figure}[tbp]
\includegraphics[width=2in,angle=-90]{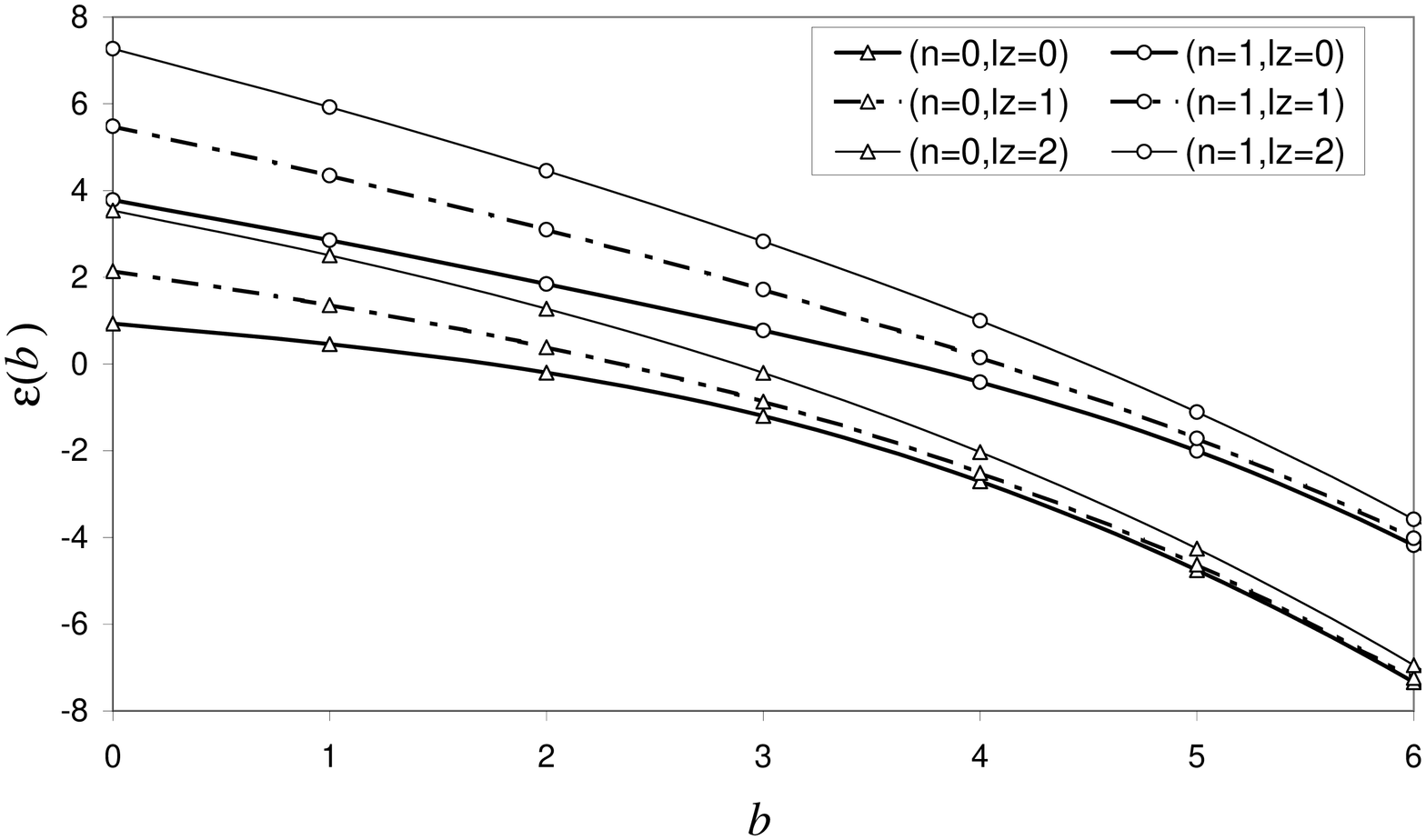}
\caption{Dependencies of several lowest eigenvalues
$\varepsilon_{n,l_{z}}$ (with $n=0,1$ and $l_{z}=0,1,2$) of
Eq.(\ref{Hpsi}) on parameter $b$.} \label{FigE}
\end{figure}

Thus, the wave function $\psi_{n,l_z}(\mathbf{q})$ describes a
particle with the azimuthal quantum number $l_z$ in the 2D axially
symmetrical potential well $U(q)=q^{4}-b q^{2}$ ($q=|\mathbf{q|}$).
Obviously, the eigenfunctions and the eigenvalues of Eq.(\ref{Hpsi})
are real for all values of $b $. The dependencies of several lowest
eigenvalues of Eq.(\ref{Hpsi}) on model parameter $b$ are shown in
Fig.\ref{FigE}. For large $b $ the potential $U(q)$ takes the form
of the Mexican hat with a deep and steep axially symmetrical well
$U(q)\approx 2b (q-\sqrt{b/2})^2-b^2/4$. In this case the spectrum
becomes almost degenerate over angular momentum $l_z$, which is in
accord with Fig.\ref{FigE}. In the limit $b\to\infty$ the asymptotic
of eigenvalues of Eq.(\ref{Hpsi}) can be found explicitly:
\begin{equation}
\varepsilon_{n,l_z}=-\frac{b^2}{4}+n\sqrt{2b } +\frac{l_z^2}{4b}
\label{epsilon}
\end{equation}

The partition function is a sum of exponents over all states and in
the thermodynamic limit ($\lambda =2T^{1/3}L\to\infty$) only the
lowest eigenvalue $\varepsilon_0$ with $n=l_z=0$ gives contribution
to the partition function:
\begin{equation}
Z=\sum_{n,l_z}e^{-\lambda \varepsilon _{n,l_{z}}}\to e^{-\lambda
\varepsilon_0}  \label{z3}
\end{equation}

\section{Correlation function}

The calculation of the correlation function $\left\langle
\vec{s}(l)\cdot \vec{s}(0)\right\rangle $ is more complicated
procedure in comparison with that for the partition function. It is
considered in detail for the case $\alpha =1/4$ ($b =0$) in
Ref.\cite{DK102}, where the problem of the calculation of the
correlation functions reduces to solution of the additional
differential equations together with Eq.(\ref{Hpsi}). This procedure
can be extended to the case $\alpha \neq 1/4$ straightforwardly. In
particular, the expressions for the correlation function remains
exactly the same as for the case $\alpha =1/4$ and only the form of
the potential well in the corresponding pair of differential
equations is modified, $U(\mathbf{q})=\mathbf{q}^{4}-b
\mathbf{q}^{2}$ instead of $U(\mathbf{q})=\mathbf{q}^{4}$. The
differential equations are
\begin{eqnarray}
&&-\frac{1}{4}\frac{d^{2}v}{dq^{2}}-\frac{1}{4q}\frac{dv}{dq}
+\frac{1}{4q^{2}}v+q^{4}v-b q^{2}v+qu=\eta v  \nonumber \\
&&-\frac{1}{4}\frac{d^{2}u}{dq^{2}}-\frac{1}{4q}\frac{du}{dq}+q^{4}u-b
q^{2}u-qv=\eta u  \label{ode2}
\end{eqnarray}

Eqs.(\ref{ode2}) represent the eigenvalue problem for $\eta_n$ and
eigenfunctions $u_n(q)$ and $v_n(q)$ satisfying the normalization
conditions:
\begin{equation}
\left\langle u_{n}^{\ast }|u_{m}\right\rangle -\left\langle v_{n}^{\ast
}|v_{m}\right\rangle =\delta _{n,m}  \label{norma3d}
\end{equation}

System of equations (\ref{ode2}) describes a two-level system in the
axially symmetric potential well $U(\mathbf{q})=\mathbf{q}^{4}-b
\mathbf{q}^{2}$, where two levels with angular momenta $l_z=0$ and
$l_z=1$ are coupled by non-Hermitian transition operator.

As shown in Ref.\cite{DK102} the spin correlation function has a
form
\begin{equation}
\left\langle \vec{s}(l)\cdot \vec{s}(0)\right\rangle =\sum_{n}\left\langle
\psi _{0}|u_{n}\right\rangle ^{2}e^{-2T^{1/3}(\eta _{n}-\varepsilon _{0})l}
\label{cor}
\end{equation}
where $\psi_0$ and $\varepsilon_0$ are the ground state wave
function and the ground state energy of the Schr\"{o}dinger equation
(\ref{Hpsi}) and $u_n$ and $\eta_n$ are the eigenfunctions and the
eigenvalues of Eqs.(\ref{ode2}).

\begin{figure}[tbp]
\includegraphics[width=2in,angle=-90]{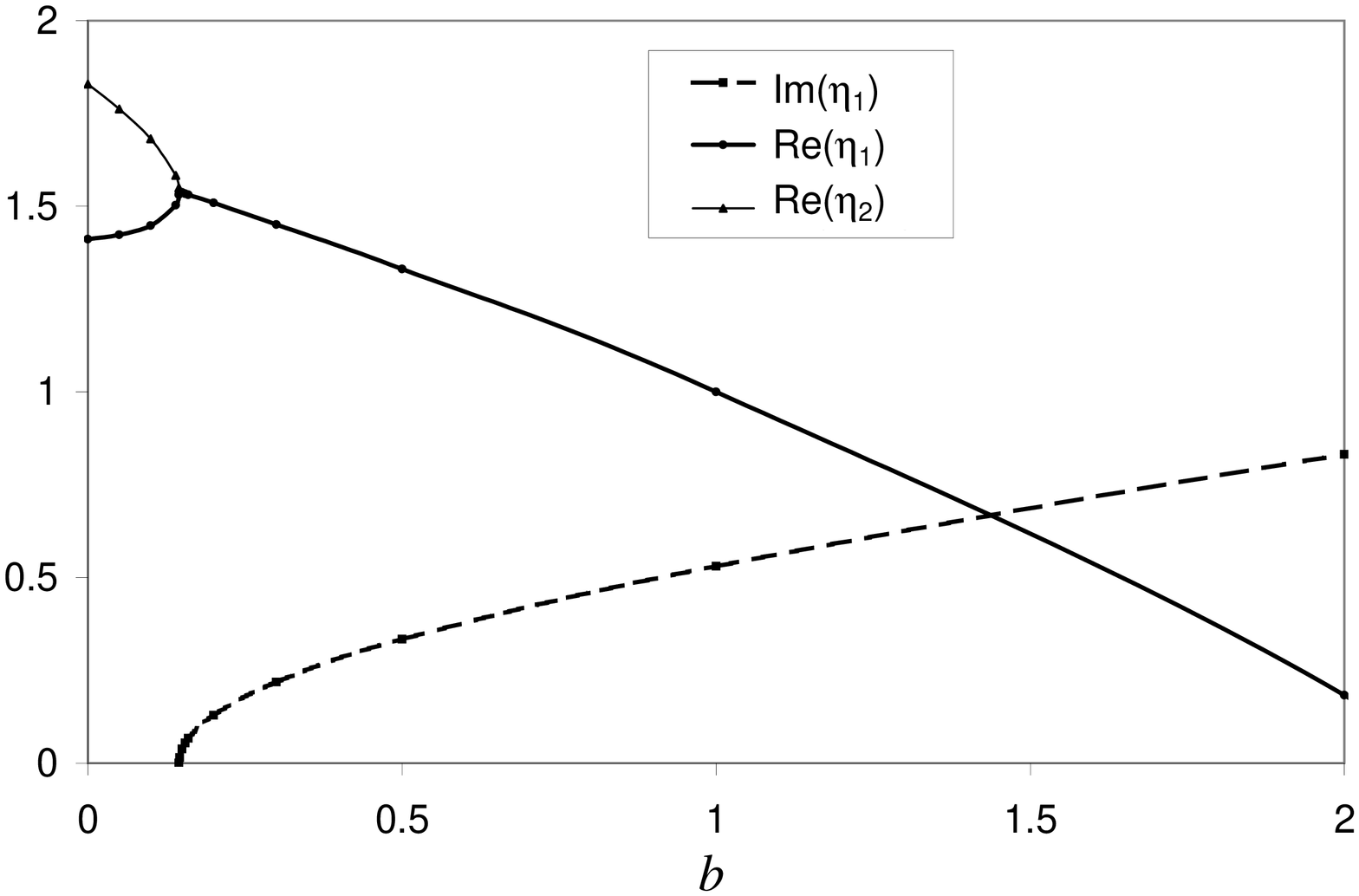}
\caption{Dependencies of eigenvalues $\eta_1$ and $\eta_2$ on
parameter $b$. Eigenvalues $\eta_1$ and $\eta_2$ are real for
$b<0.145$: real part of $\eta_1$ (thick solid line) and $\eta_2$
(thin solid line). Eigenvalues $\eta_1$ and $\eta_2$ are complex
conjugated at $b>0.145$: $\Re(\eta_1)=\Re(\eta_2)$ (thick solid
line) and $\Im(\eta_1)=-\Im(\eta_2)$ (dashed line).} \label{Figeta}
\end{figure}

The numerical solution of Eqs.(\ref{ode2}) demonstrates the
following properties. For small $b$ all eigenvalues of
Eq.(\ref{ode2}) are real but the eigenfunctions of Eq.(\ref{ode2})
are complex. With the increase of $b$, two lowest levels
$\eta_{1,2}$ approach to each other, then at $b_0\simeq 0.145$ they
coincide and after that, for $b>b_0$, these eigenvalues become
complex conjugated. Their imaginary part increases as
$\Im(\eta_{1,2})\sim\sqrt{(b-b_0)/2} $. These features of the
solutions of Eqs.(\ref{ode2}) are illustrated in Fig.\ref{Figeta},
where the dependencies of the eigenvalues $\eta_{1,2}$ on parameter
$b$ are shown. According to Eq.(\ref{cor}) the appearance of
imaginary parts of eigenvalues $\eta_n$ implies that the correlation
function begins to oscillate.

With the further increase of $b$ the next two levels $\eta _{3,4}$
approach to each other and at some value $b_1>b_0$ become complex
conjugated. And so on, so that at large values of $b$ all low-lying
levels are divided on complex conjugated pairs. The corresponding
eigenfunctions of these pairs are complex conjugated as well. This
fact guarantees the reality of the correlation function.

It is important that all the eigenvalues $\varepsilon_0$, $\eta_n$
and the eigenfunctions $\psi_0$, $u_n$ presented in Eq.(\ref{cor})
depend on the parameter $b$ only. This means that the correlation
function and other related physical quantities are universal
functions of the scaling parameter $b$.

In what follows it is convenient to consider the system properties
as functions of the scaling parameter $t=b^{-3/2}$, which represents
the normalized temperature $t=T/\gamma^{3/2}$. As shown in the
Appendix the excitation energy $E_{\mathrm{dw}}$ of a chiral domain
wall, which separates two domains with different chirality vector,
is proportional to $E_{\mathrm{dw}}\sim\gamma^{3/2}$. Therefore, the
parameter $t$ is nothing but the ratio (up to numerical factor)
$T/E_{\mathrm{dw}}$ and the physical meaning of the scaling variable
$t$ is the normalization of the temperature by the energy of the
chiral domain wall.

According to Eq.(\ref{cor}) the static structure factor is
\begin{equation}
S(k)=\frac{1}{T^{1/3}}\sum_{n}\left\langle \psi_0|u_n\right\rangle^2
\frac{\eta_n-\varepsilon_0} {(\eta_n-\varepsilon_0)^2 +\tilde{k}^2}
\label{Sk2d}
\end{equation}
where $\tilde{k}=k/2T^{1/3}$ is the scaled momentum. As follows from
Eq.(\ref {Sk2d}), the structure factor is scaled as $T^{-1/3}$ and,
therefore, $ S(k)T^{1/3}$ is the universal function of $t$.

\begin{figure}[tbp]
\includegraphics[width=2in,angle=-90]{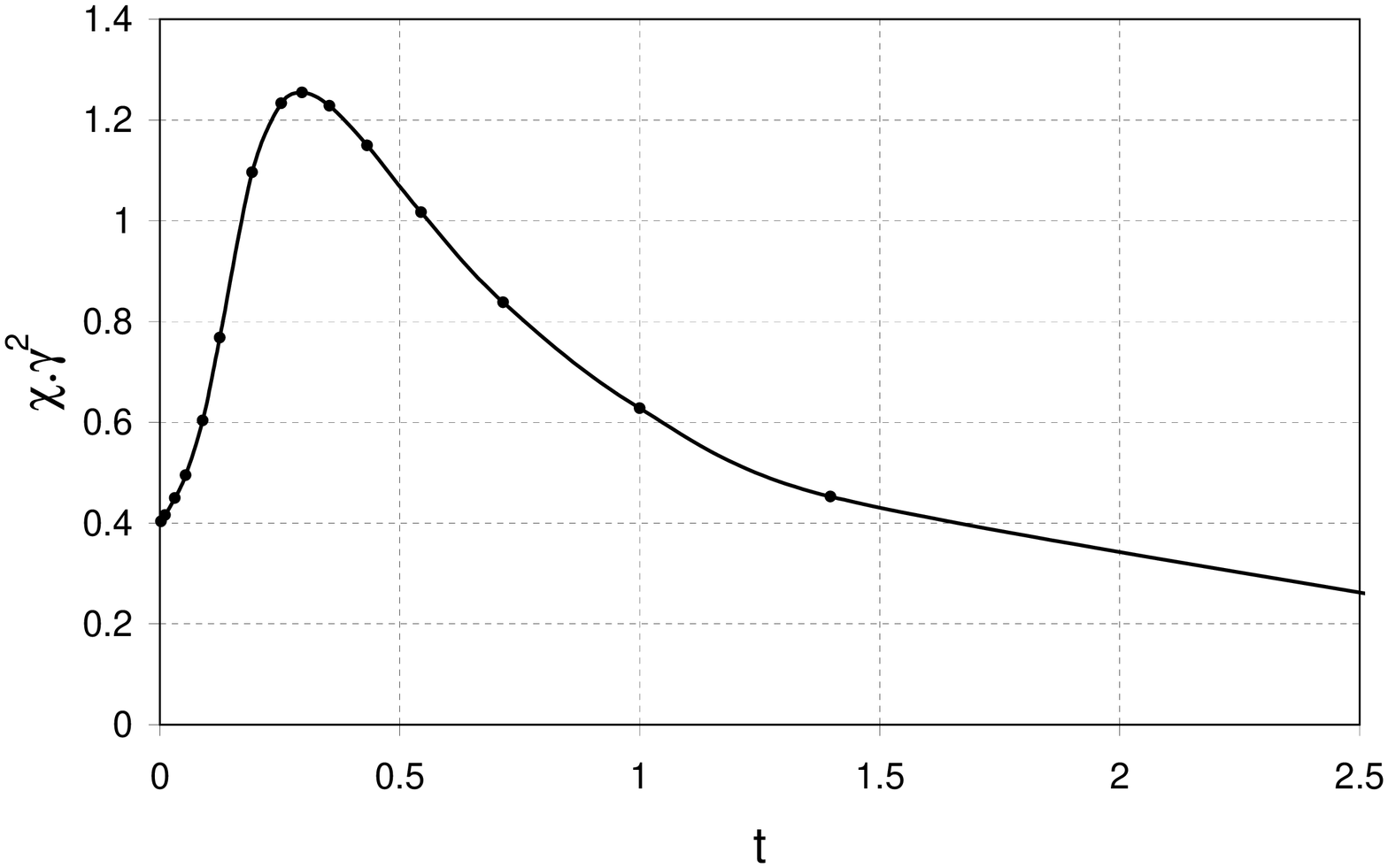}
\caption{The dependence of the normalized susceptibility
$\tilde{\chi }=\chi\gamma^2$ on the scaled temperature $t$.}
\label{Figchi}
\end{figure}

The susceptibility $\chi(k)$ is connected with the static structure
factor by the relation $\chi(k)=S(k)/3T$. At first, we consider the
temperature dependence of the uniform susceptibility $\chi(0)$.
According to Eq.(\ref{Sk2d}) it is
\begin{equation}
\chi(0)=\frac{1}{3T^{4/3}}\sum_{n}\frac{\left\langle \psi
_{0}|u_{n}\right\rangle^2}{\eta_n-\varepsilon_0} \label{chi0}
\end{equation}

Since the eigenfunctions and the eigenvalues of Eqs.(\ref{Hpsi}) and
(\ref{ode2}) are functions of a single variable $t$, the normalized
susceptibility $\tilde{\chi}=\chi\gamma^2$ can be expressed as an
universal function of the normalized temperature $t$. Thus,
$\tilde{\chi}$ reveals the scaling behavior: in the vicinity of the
transition point and at $T\to 0$ the dependence of $\tilde{\chi}$ on
the temperature and the model parameter $\gamma $ enters only in the
combination $T/\gamma^{3/2}$. The dependence of $\tilde{\chi}$ on
$t$ is shown in Fig.\ref{Figchi}. The characteristic features of
this dependence are the existence of maximum of $\tilde{\chi}$ at
$t=t_{m}$ ($t_{m}\simeq 0.432$) and the finite value of
$\tilde{\chi}$ at $T=0$. The latter fact is a pure classical effect.
In the $s=1/2$ quantum F-AF model the ground state is believed to be
gapped (though the gap can be extremely small \cite{Itoi}), and so
$\tilde{\chi}\to 0$ as $T\to 0$. The obtained dependence
$\tilde{\chi}(t)$ means that for a given model parameter $\gamma$
the uniform susceptibility has the maximum at temperature $T_{m}\sim
\gamma^{3/2}$ and the height of this maximum is $\chi_{m}\sim \gamma
^{-2}$. This implies that with the increase of $\gamma$ the maximum
of $\chi$ shifts to higher temperatures and the magnitude of the
maximum $\chi_m$ decreases.

\begin{figure}[tbp]
\includegraphics[width=2in,angle=-90]{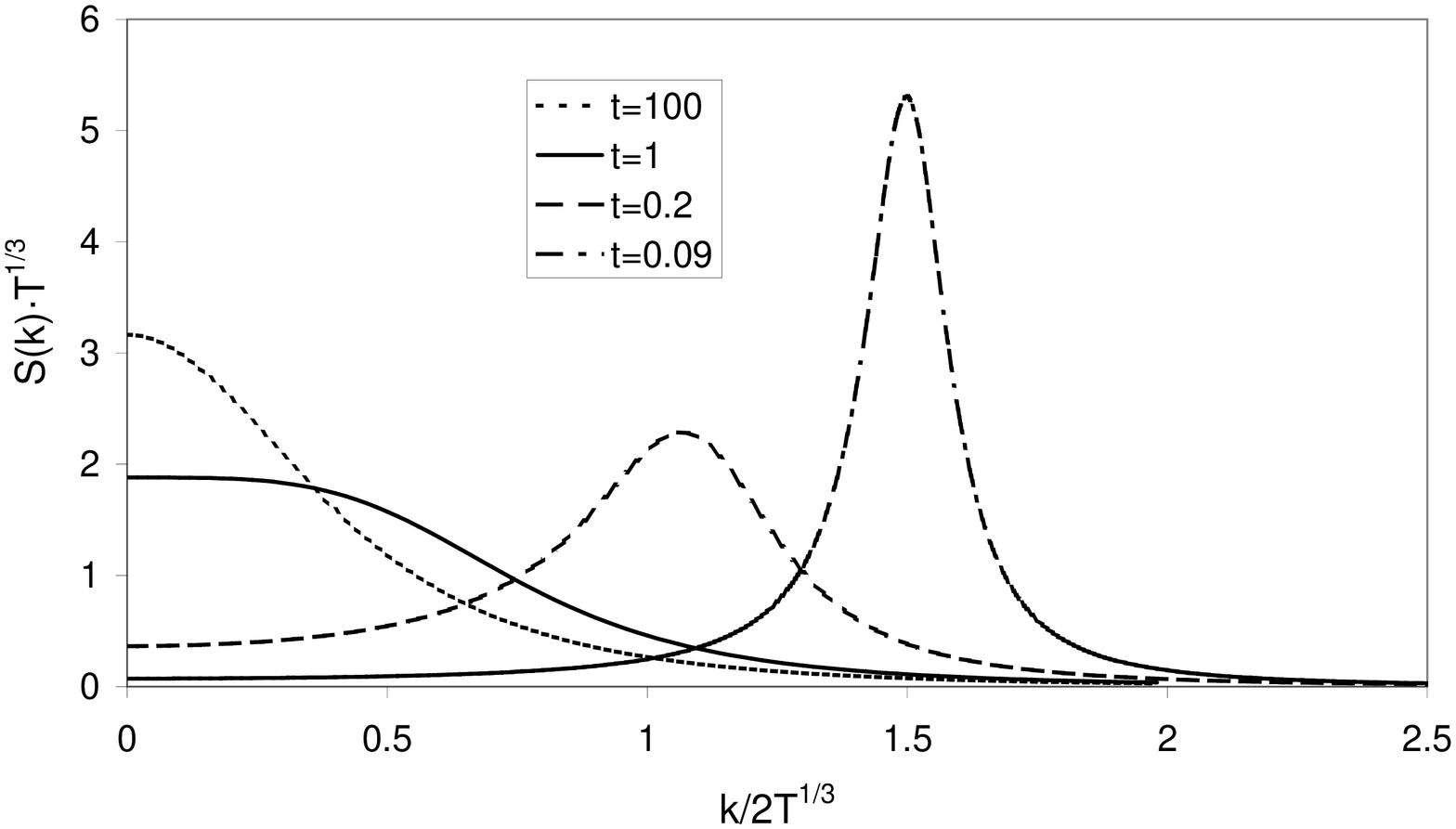}
\caption{The static structure factor $S(k)$ for several values of
the scaled temperature $t$.} \label{FigSk}
\end{figure}

Let us consider the behavior of the static structure factor
(\ref{Sk2d}) as a function of $k$. Some examples of $S(k)$ are shown
in Fig.\ref{FigSk}. As we see, the static structure factor has a
pronounced maximum and its location depends on $t$. In the vicinity
of the maximum $S(k)$ can be represented as
\begin{equation}
S(k)=\frac{S(k_{m})}{1+l_{c}^{2}(k-k_{m})^{2}}
\end{equation}

Therefore, the behavior of the static structure factor is
described by three parameters: the location of the maximum
$k_{m}$, the magnitude of the maximum $S(k_{m})$, and the
correlation length $l_{c}$ characterizing the width of the
maximum:
\begin{equation}
l_{c}^{2}=-\frac{1}{2S(k_{m})}\frac{d^{2}S(k_{m})}{dk^{2}}
\end{equation}

Obviously, the correlation length is scaled as the space variable
(\ref{xi}) and, therefore, the normalized correlation length
$\tilde{l}_c=2T^{1/3}l_{c}$ is the universal function of $t$.

The location of the maximum $k_{m}$ is determined by the condition
$dS/dk=0$. According to Eq.(\ref{Sk2d}) this condition reduces to
the equation
\begin{equation}
\tilde{k}_{m}\sum_{n}\frac{\left\langle \psi
_{0}|u_{n}\right\rangle ^{2}(\eta _{n}-\varepsilon _{0})}{[(\eta
_{n}-\varepsilon _{0})^{2}+\tilde{k} _{m}^{2}]^{2}}=0
\label{eqkm}
\end{equation}

This equation determines the scaled momentum $\tilde{k}_{m}$ as a
function of $t$. Obviously, Eq.(\ref{eqkm}) always has a trivial
solution $\tilde{k}_{m}=0$. For high values of $t$ there are no
other solutions. However, at some value of the scaled temperature
$t=t_{c}$ another, non-trivial solution of Eq.(\ref{eqkm}) with
$\tilde{k}_{m}\neq 0$ appears. For $t<t_{c}$ the maximum of the
structure factor shifts from $\tilde{k}_{m}=0$. Thus, $t=t_{c}$
defines the Lifshitz point which corresponds to the transition from
the helical to the ferromagnetic phase. Certainly, there is no true
helical or ferromagnetic LRO in these phases, because thermal
fluctuations destroy any LRO in one-dimensional systems. Therefore,
under the helical and the ferromagnetic phases we mean the presence
of the short-range order (SRO) of the corresponding type.

The location of the Lifshitz point is determined by the equation:
\begin{equation}
\sum_{n}\frac{\left\langle \psi _{0}|u_{n}\right\rangle ^{2}}{(\eta
_{n}-\varepsilon _{0})^{3}}=0  \label{eqLifschitz}
\end{equation}

\begin{figure}[tbp]
\includegraphics[width=2in,angle=-90]{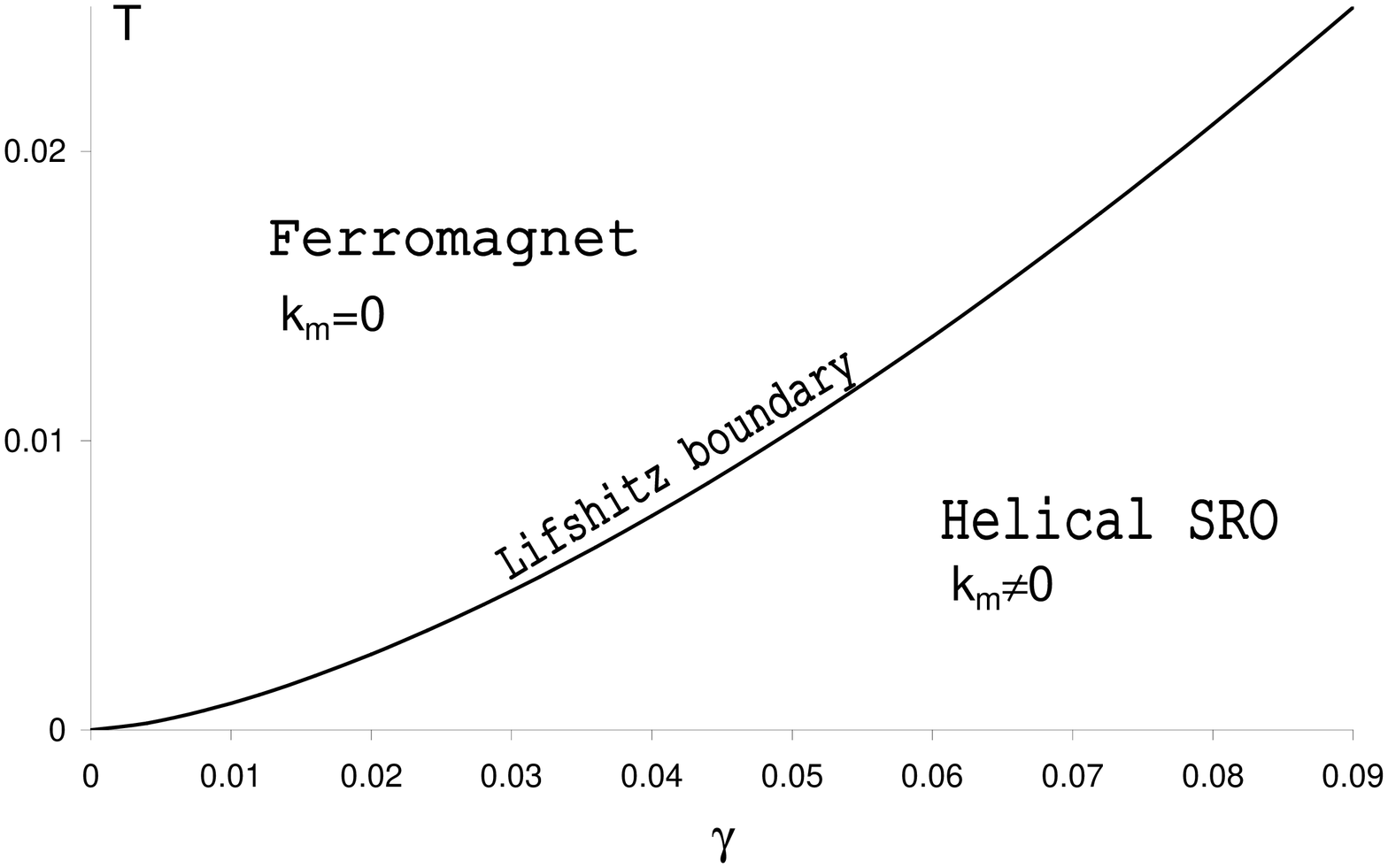}
\caption{The phase diagram of model (\ref{Etr}).} \label{Figphase}
\end{figure}

The numerical solution of Eq.(\ref{eqLifschitz}) gives
$t_{c}\simeq 0.925$. Thus, the transition line between the
ferromagnetic and the helical phases (`Lifshitz boundary') in
($T,\gamma$) plane has a form
\begin{equation}
T_{c}(\gamma )=0.925\gamma ^{3/2}
\end{equation}

The corresponding phase diagram is shown in Fig.\ref{Figphase}.

\begin{figure}[tbp]
\includegraphics[width=2in,angle=-90]{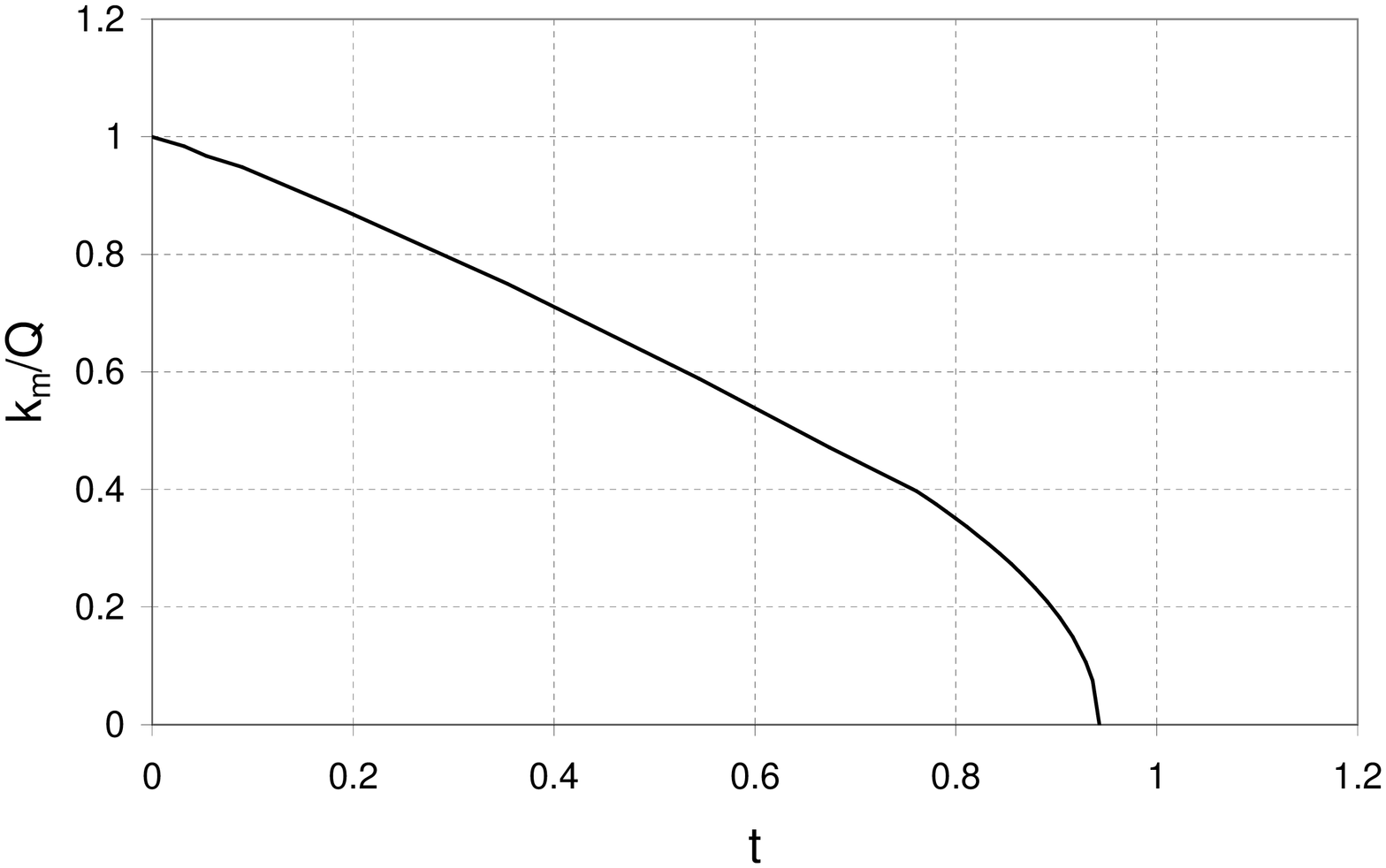}
\caption{The dependence of $k_{m}/Q$ ($Q=\sqrt{2\gamma}$) on the
scaled temperature $t$.} \label{Figkmax}
\end{figure}

\begin{figure}[tbp]
\includegraphics[width=2in,angle=-90]{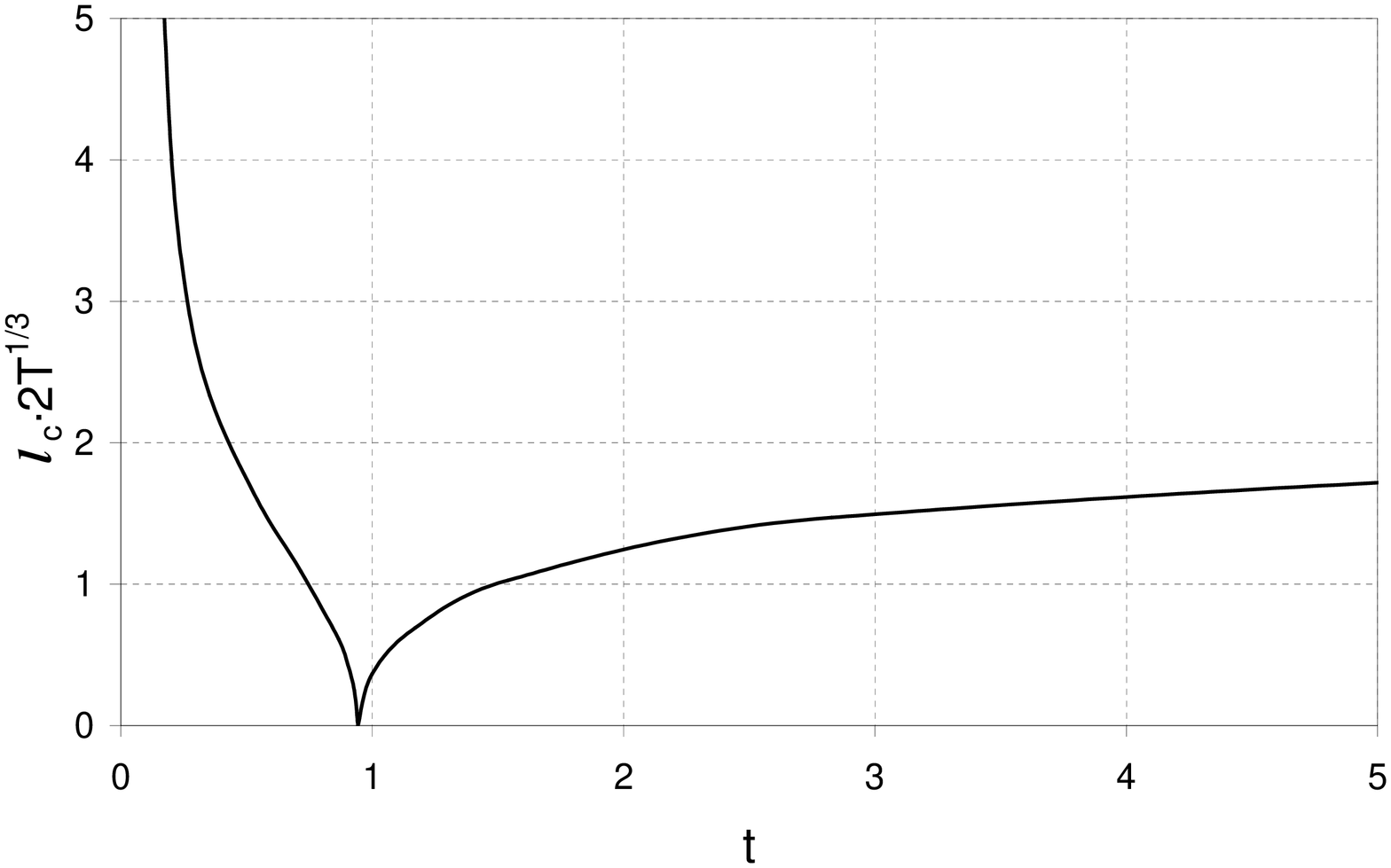}
\caption{The dependence of the normalized correlation length
$\tilde{l}_c=2T^{1/3}l_{c}$ on the scaled temperature $t$.}
\label{Figlc}
\end{figure}

For a fixed model parameter $\gamma \ll 1$ the dependence of the
system properties on the temperature can be qualitatively described
in the following way. At $T=0$ the structure factor $S(k)$ has a
$\delta $-function form at the corresponding wave vectors
$k_{m}=\pm\sqrt{2\gamma}$, indicating the helical LRO. At low
temperatures ($T\ll T_{c}$) the $\delta $-peaks are smeared by
gapless excitations (spin waves) over the helical ground states. The
chiral domain wall excitations have the gap ($E_{\mathrm{dw}}\sim
\gamma ^{3/2}$) and their contribution to the thermodynamics is
exponentially small. The analysis of Eq.(\ref{Sk2d}) shows that in
this region ($T\ll T_{c}$) the magnitude of the maximum and the
correlation length behave as $S(k_{m})\sim l_{c}\sim \gamma /T$.

When the temperature is increased, the maximum of $S(k)$ shifts to
$k=0$ and at approaching to the Lifshitz point the value of $k_{m}$
tends to zero by the square root law $k_{m}\sim \gamma
^{-1/4}\sqrt{T_c-T}$. This is similar to the order parameter
behavior in Landau theory of second-order phase transitions. Indeed,
in the three-dimensional frustrated classical spin model, when the
helical LRO exists, the wave vector associated with the helical
phase vanishes by the same law \cite{Redner}.

The numerically obtained dependencies of $k_{m}/\sqrt{2\gamma}$ and
the scaled correlation length $\tilde{l}_{c}$ on $t$ are
demonstrated in Fig.\ref{Figkmax} and Fig.\ref{Figlc},
correspondingly. As shown in Fig.\ref{Figlc} the correlation length
vanishes at $t=t_{c}$, because $S(k)$ does not contain $k^{2}$ term
in $k$ expansion at the Lifshitz point. The magnitude of the maximum
at the Lifshitz point is $S(0)\approx 1.9\gamma ^{-1/2}$.

For $T\gg T_{c}$ the helical correlations are destroyed, the maximum
of $S(k)$ is situated at $k_{m}=0$ and in the limit $t\to\infty $ it
tends to the results obtained in Ref.\cite{DK102}:
$S(0)=3.21/T^{1/3}$ and $l_{c}=1.04/T^{1/3}$.

Up to now we considered the classical spin model where $\vec{S}$
is usual three-component classical vector. It is appropriate to
discuss briefly the planar model describing by Hamiltonian
(\ref{H14}), where the vector $\vec{S} $ has two components
(planar spin model). For short we denote below these two models as
$\nu =3$ and $\nu =2$ cases. Just as for the $\nu =3$ model the
transition point between the ferromagnetic and the helical ground
states for the $\nu =2$ model takes place at $\alpha =1/4$. The
low-temperature thermodynamics of the $\nu =2$ model near the
transition point can be analyzed in a similar manner as for the
$\nu =3$ case. In particular, the calculation of the partition
function for the $\nu =2$ model reduces to the solution of the
Schr\"{o}dinger equation
\begin{equation}
-\frac{1}{4}\frac{d^{2}\psi }{dq^{2}}+U(q)\psi =\varepsilon \psi
\label{schr}
\end{equation}
where $U(q)=q^{4}-b q^{2}$.

Our study of the planar model shows that the thermodynamic
quantities are the universal functions of the same scaling parameter
$t=T/\gamma^{3/2}$. The behaviors of $S(k)$, $k_{m}(t)$ and
$l_{c}(t)$ are similar to that for $\nu=3$ case. The Lifshitz point
is determined by $t_{c}\simeq 0.7$ and the critical exponent
characterizing the behavior of $k_{m}$ near $t_{c}$ is $1/2$ as
well. However, the behavior of the chiral correlation function
\begin{equation}
R_{l}=\left\langle \vec{K}_{n}\cdot \vec{K}_{n+l}\right\rangle
\end{equation}
of these two models are different. In the continuum approach the chiral
vector is expressed as
\begin{equation}
\vec{K}_{n}=\vec{S}_{n}\times \vec{S}_{n+1}\approx
\vec{s}(x)\times \frac{
\partial \vec{s}(x)}{\partial x}
\end{equation}

In the $\nu=2$ model the chiral vector is directed perpendicular to
the spin plane for any $x$. Therefore in this case
\begin{equation}
\vec{K}(l)\cdot \vec{K}(0)\approx \frac{\partial
\vec{s}(l)}{\partial x} \cdot \frac{\partial \vec{s}(0)}{\partial
x}=q(l)q(0)
\end{equation}

In the $\nu=3$ model we take into account local coordinates with
$\vec{s}=(0,0,1)$ and $q_z=0$ and obtain similar expression
\begin{equation}
\vec{K}(l)\cdot \vec{K}(0)=\mathbf{q}(l)\cdot \mathbf{q}(0)
\end{equation}

The expression for the chiral correlation function $R_l$ can be
obtained in the same manner as for the spin correlation function:
\begin{equation}
R_l = \sum_{n}\left\langle \psi_0|\mathbf{q}|\psi_n\right\rangle
^{2} e^{-2T^{1/3}(\varepsilon_n-\varepsilon_0)l} \label{R}
\end{equation}

Here $\psi_n$ and $\varepsilon_n$ are the eigenfunctions and the
eigenvalues of the Schr\"{o}dinger equations (\ref{Hpsi}). For
planar case $R_l$ has similar expression where $\psi_n$ and
$\varepsilon_n$ are solutions of Eq.(\ref{schr}). Since $\psi_n$ and
$\varepsilon_n$ are always real, the chiral correlation function
$R_l$ does not oscillate and the maximum of the corresponding
structure factor is at $k=0$ for both $\nu=2$ and $\nu=3$ cases.

As follows from Eq.(\ref{R}) the correlation length of the chirality
$\zeta$ is scaled as $T^{1/3}$ and at $T\to 0$ it is defined by the
lowest level having the non-zero matrix element with the
ground-state wave function $\psi_0$. So, for normalized chiral
correlation length $\tilde{\zeta}=2T^{1/3}\zeta$ we have
\begin{equation}
\tilde{\zeta}=\frac{1}{\varepsilon_1-\varepsilon_0}
\end{equation}

The ground state wave function for Eq.(\ref{schr}) is an even
function of $q$ and only the eigenfunctions with odd parity give
contributions to $R_l$. So, for the $\nu=2$ model $\varepsilon_1$ is
the energy of the first excited state of the odd parity. For the
$\nu=3$ model the matrix elements in Eq.(\ref{R}) are non-zero only
for the wave functions $\psi_n$ with $l_z=1$. At the transition
point (corresponding to the limit $t\to\infty$) the energy
difference $\varepsilon_1-\varepsilon_0$ is finite for both $\nu=2$
($\varepsilon_1-\varepsilon_0=1.09$) and $\nu=3$
($\varepsilon_1-\varepsilon_0=1.21$) cases.

At $t\to 0$ the behavior of the energy gap between two lowest levels
for $\nu=2$ and $\nu=3$ models becomes qualitatively different. For
the $\nu=2$ case the potential $U(q)=q^4-b q^2$ has two degenerate
minima and the splitting of two lowest levels
$\varepsilon_1-\varepsilon_0$ depends upon tunneling through the
classically forbidden region of small $q$. This tunneling is
exponentially small for low temperatures and the WKB estimate gives
$\tilde{\zeta}\sim \exp(1/3\sqrt{2}t)$. However, for the $\nu=3$
case the particle moves in two-dimensional axially symmetric
potential, the form of the Mexican hat, and relevant splitting is
equal to the additional energy associated with one unit of the
angular momentum. Then, according to Eq.(\ref{epsilon})
$\tilde{\zeta}=4/t^{2/3}$. Thus, the behavior of the correlation
length of the chirality is essentially different for these two
models.

So far we studied the helical side of the transition point
$\alpha>1/4$ ($\gamma>0$). On the ferromagnetic side ($\gamma<0$)
the static structure factor $S(k)$ and the susceptibility $\chi(k)$
have the maximum at $k=0$ for any temperature. The uniform
susceptibility $\chi$ diverges at $T\to 0$ but the corresponding
exponent changes from $4/3$ to $2$ and Eq.(\ref{chi0}) describes
this crossover. The exponent $2$ appears in the limit $b\to
-\infty$, when $\widetilde{\chi}(b)= -2b^3/3$ and the susceptibility
becomes $\chi=2|\gamma|/3T^2$, which is in accord with the result
for HF with the coupling constant $|\gamma|$. Indeed, in the limit
$b\to -\infty$ one can neglect the second derivative term in energy
functional (\ref{Etr}) and the model reduces to the HF with
renormalized coupling $|\gamma|$.

\section{Summary}

We have studied the low-temperature thermodynamics of the
classical F-AF model in the vicinity of the ferromagnet-helimagnet
transition point with use of the continuum approximation. The
calculation of the partition and the spin correlation functions is
reduced to a quantum mechanical problem of a particle in a
potential well. It is shown that the thermodynamic quantities are
universal functions of the scaling variable $t=T/\gamma^{3/2}$,
where $\gamma=4J_2/|J_1|-1$ describes the deviation from the
transition point.

On the ferromagnetic side ($\gamma<0$) the static structure factor
$S(k)$ has the maximum at $k=0$ for all temperatures. The uniform
susceptibility $\chi$ diverges at $T\to 0$ but the corresponding
asymptotics changes from $\chi\sim T^{-4/3}$ at the transition point
to $\chi\sim |\gamma|/T^2$ for $T\ll |\gamma|^{3/2}$. The crossover
between these regimes takes place at $T\sim\gamma^{3/2}$.

On the helical side of the transition point ($0<\gamma\ll 1$) our
investigation displays the following dependence of the system
properties on the temperature. At zero temperature the static
structure factor $S(k)$ has a $\delta$-function form at the
corresponding wave vectors $k_{m}=\pm\sqrt{2\gamma}$, indicating the
LRO of the helical-type. At finite temperatures the LRO is destroyed
by thermal fluctuations, which is manifested in smearing of
$\delta$-peaks of the static structure factor by gapless excitations
(spin waves) over the helical ground states. The chiral domain wall
excitations have the gap ($E_{\mathrm{dw}}\sim\gamma^{3/2}$) and
their contribution to the thermodynamics is exponentially small for
low temperatures $T\ll \gamma^{3/2}$. In this region the magnitude
of the maximum of the structure factor and the correlation length
behave as $S(k_m)\sim l_c\sim\gamma /T$.

Further increase of the temperature is accompanied by the damping of
the peak and its shift to $k=0$. At approaching to the Lifshitz
point $T\to T_c=0.925\gamma^{3/2}$ the value of $k_{m}$ tends to
zero by the square root law $k_{m}\sim \gamma ^{-1/4}\sqrt{T_c-T}$.
Such dependence is similar to the order parameter behavior in the
Landau theory of the second-order phase transitions. So, the
location of the peak $k_{m}$ plays a role of the order parameter,
though one should remember that the system has only the short-range
helical order at finite temperatures. Equation
$T_c=0.925\gamma^{3/2}$ determines the Lifshitz boundary in
($T,\gamma$) plane, separating the ferromagnetic and the helical
phases.

For $T>T_c$ the helical correlations are destroyed by the chiral
domain wall excitations and the maximum of the static structure
factor $S(k)$ is situated at $k_{m}=0$. The high temperature limit
$T\gg T_c$ is equivalent to the limit $\gamma\ll T^{2/3}$. This
implies that one can neglect the second terms in Eqs.(\ref{H14})
and (\ref{Etr}) and the model effectively reduces to that at the
transition point studied in Ref.\cite{DK102}. Here the uniform
susceptibility and the correlation length are
$\chi(0)=1.07/T^{4/3}$ and $l_{c}=1.04/T^{1/3}$. We should remind
that our analysis is restricted by low temperatures and,
therefore, the latter high-$T$ limit is in fact restricted by
$T_c\ll T\ll 1$.

The uniform susceptibility $\chi$ has the maximum at $T_{m}\sim
\gamma^{3/2}$ and $\chi_m\sim\gamma^{-2}$, i.e. with the increase of
the frustration parameter $\alpha$ the maximum of $\chi$ shifts to
higher temperatures and the magnitude of the maximum $\chi_m$
decreases. The obtained dependence of $\chi$ on $T$ is in a
qualitative agreement with those observed in the edge-shared
compounds with $\alpha$ close to $1/4$ ($Li_{2}CuO_{2}$ \cite
{Mizuno}, $Rb_{2}Cu_{2}Mo_{3}O_{12}$ \cite{Hase}, $Li_{2}CuZrO_{4}$
\cite {Volkova}). The dependencies of $T_m$ and $\chi_m$ on the
frustration parameter $\alpha$ are also in accord with the
experimental observations \cite{Volkova}.

We have also shown that the thermodynamic behavior of both planar
and classical F-AF models are similar. The exception is the chiral
correlation function. The chiral correlation length of the planar
model increases exponentially at $T\to 0$ in contrast with the power
dependence for the classical $\nu=3$ model. This fact is explained
by different topology of the potentials in the corresponding
Schr\"{o}dinger equations.

It is interesting to compare our results with those obtained in
Refs.\cite{Harada0,Harada} for intermediate values of $T$ and
$\gamma$. In contrast with the phase diagram obtained for small
values of $T$ and $\gamma$ and schematically shown in Fig.6, the
Lifshitz boundary for intermediate values of $T$ and $\gamma$ can
have the reentrant behavior \cite{Harada0}. There is a critical
value of $\gamma=\gamma_c$ (for the planar model $\gamma_c\simeq
0.16$ \cite{Harada0}), where the Lifshitz boundary turns round and
moves back to $\gamma=0$ with the further increase of temperature.
This means that for a fixed small value of $\gamma$
($0<\gamma<\gamma_c$) the ferromagnetic phase exists in the finite
range of temperatures and for $\gamma>\gamma_c$ the helical-type
correlations remain for all temperatures. This fact testifies that
in the region of intermediate or large values of $T$ and $\gamma$,
where the continuum approximation becomes unapplicable, the
thermodynamic behavior can be qualitatively different from the
results presented here for small $T$ and $\gamma$.

\appendix*\section{}

In the appendix we calculate the chiral domain wall energy for the
classical spin model described by energy functional (\ref{Etr}).
To study excitations in this model it is useful to make the
following rescaling of the space variable $x=\xi /\sqrt{2\gamma}$.
Then, energy functional (\ref{Etr}) takes the form:
\begin{equation}
E=\frac{\gamma ^{3/2}}{2^{3/2}}\int d\xi \left[ \left( \frac{\partial^2
\vec{s}}{\partial \xi^2}\right)^2 -2\left( \frac{\partial \vec{s}}
{\partial \xi }\right)^2 \right] \label{Esxi}
\end{equation}

Here we see that the integrand in Eq.(\ref{Esxi}) does not contain
any parameter and, therefore, any localized excitation is
proportional to $\sim\gamma^{3/2}$. Certainly, the determination
of the numerical factor before $\gamma^{3/2}$ needs the solution
of the corresponding Euler equation. These Euler equations are
highly non-linear and its complete analysis is hardly possible.
Fortunately, such an analysis is possible for the planar spin
case. The solutions for the planar spin case represent particular
cases of solutions for the original three-component spin model.

For the planar model the two-component spin vector can be
represented as $\vec{s}=(\sin\theta,\cos\theta)$ and energy
functional (\ref{Esxi}) in terms of $\theta(x)$ has a form
\begin{equation}
E=\frac{\gamma ^{3/2}}{2^{3/2}}\int d\xi \left[ \left(
\frac{\partial^2\theta }{\partial \xi^2}\right)^2 +\left( \frac{\partial \theta }
{\partial \xi }\right)^4 -2\left( \frac{\partial \theta }{\partial \xi }
\right)^2 \right]   \label{Exi}
\end{equation}

Variation of energy functional (\ref{Exi}) in $\theta (\xi)$ leads
to the Euler equation
\begin{equation}
\frac{\partial^4\theta }{\partial \xi^4}-6\frac{\partial^2\theta }
{\partial \xi^2}\left( \frac{\partial \theta }{\partial \xi }\right)^2
+2\frac{\partial^2\theta }{\partial \xi^2}=0  \label{euler}
\end{equation}

Here we see that the Euler equation has a trivial helical solution
$\theta (\xi )=a(\xi -\xi_0)$ with any constants $a$ and $\xi_0$.
However, the minimum of energy (\ref{Exi}) corresponding to the
ground state is achieved by a definite helical configuration with
\begin{equation}
\theta (\xi )=\pm (\xi -\xi_0)  \label{spiral}
\end{equation}
and the ground state energy is
\begin{equation}
E_{0}=-L\frac{\gamma^2}{2}  \label{Emin}
\end{equation}

To find the excitations over the helical ground state we should
look for other solutions of Eq.(\ref{euler}). Eq.(\ref{euler}) can
be integrated so that for $y(\xi )=\partial \theta /\partial \xi $
we obtain equation
\begin{equation}
y^{\prime 2}=y^{4}-2y^{2}+Ay+B  \label{ode}
\end{equation}
with some integration constants $A$ and $B$.

The energy functional in terms of $y(\xi)$ takes the form:
\begin{equation}
E-E_0=\frac{\gamma ^{3/2}}{2^{3/2}}\int \mathrm{d}\xi
\left[ y^{\prime 2}+(1-y^{2})^{2}\right]  \label{Ey}
\end{equation}

The right-hand side of Eq.(\ref{ode}) has a double-well form. The
complete analysis of solutions of Eq.(\ref{ode}) will be given
elsewhere. Here we present the solution describing the chiral
domain wall only. This case corresponds to the choice of
integration constants $A=0$ and $B=1$. In this case the solution
of Eq.(\ref{ode}) has a simple form
\begin{equation}
y(\xi )=\tanh \left( \xi -\xi _{0}\right)  \label{th}
\end{equation}
or in original terms for angle $\theta (x)$
\begin{equation}
\theta(x)=\theta _{0}+\ln \left[ \cosh \left( \sqrt{2\gamma}
(x-x_{0}\right) \right]
\end{equation}
with arbitrary $\theta_0$ and $x_0$. As can be seen, this solution
describes a chiral domain wall separating two domains of opposite
chirality. The energy of this excitation is
\begin{equation}
E_{\mathrm{dw}}=\frac{1}{3}(2\gamma)^{3/2}  \label{Ekink}
\end{equation}

In case of three-component spin vectors the chiral domain wall
excitations are more complicated. They depend on the boundary
conditions on $x=\pm\infty$, which can be imposed on the chiral
vectors $\vec{K}$. In the helical ground state the length of the
chiral vectors equals $|\vec{K}|=\sqrt{2\gamma}$, but the
direction is arbitrary according to degeneracy of the helical
ground state. Therefore, the chiral domain wall solution is a
function of the angle between the chiral vectors at $x=\pm\infty$:
\begin{equation}
\cos\phi=\frac{(\vec{K}_{-\infty} \cdot \vec{K}_{\infty})}
{2\gamma}
\end{equation}

In the planar spin case considered above the angle $\phi$ is
$\phi=\pi$, because chiral vectors are antiparallel on $\pm\infty$.
It is also obvious that the limit $\phi\to 0$ corresponds to the
ground state configuration. In general the dependence of the energy
of the chiral domain wall on the angle $\phi$ is unknown, but it is
natural to assume that this function monotonically increases from
zero at $\phi=0$ to the exact result Eq.(\ref{Ekink}) at $\phi=\pi$.
Therefore, in any case the energy of the chiral domain wall is
proportional to $\sim\gamma^{3/2}$ and can be written as
\begin{equation}
E_{\mathrm{dw}}=f(\phi)\gamma^{3/2}  \label{Ekink3d}
\end{equation}
with some smooth monotonic function $f(\phi)$ having the limits
$f(0)=0$ and $f(\pi)=2^{3/2}/3$.

\end{document}